\RequirePackage{fix-cm}

 \documentclass[showkeys,preprint,nofootinbib]{revtex4}          % twocolumn%
\usepackage{graphicx}
\usepackage{amssymb}
\usepackage{amsbsy}
 \usepackage{amsmath}
\usepackage{subfigure}
\usepackage{color}
\usepackage{enumitem}
\usepackage{easybmat}

\newcommand{\modif}[1]{#1}

\begin{document}

\title[]{Computation of effective electrical conductivity of composite materials: a novel approach based
on analysis of graphs.
}

{ \hspace{0.7\linewidth} {To Vojislav Golo...}}

\author{Vladimir Salnikov} 
\email{vladimir.salnikov@univ-lr.fr}
\affiliation{LaSIE, University of La Rochelle,
{Av. Michel Cr\'epeau, 17042 La Rochelle, France }}

\author{Daniel Cho\"i}
\email{daniel.choi@unicaen.fr}
\author{Philippe Karamian-Surville}
\email{philippe.karamian@unicaen.fr}

\affiliation{%Nicolas Oresme Mathematics Laboratory, 
LMNO, University of Caen Normandy, CS 14032, Bd. Mar\'echal Juin,  BP 5186, 
  14032, Caen Cedex,  France
}


%\date{19 September 2014}

\begin{abstract}
In this work we continue the investigation of different approaches 
to conception and modeling of composite materials. 
The global method we focus on, is called 'stochastic homogenization'. In this approach, the classical deterministic \emph{homogenization} techniques and procedures are used to compute the macroscopic parameters of a composite starting from its microscopic properties. The \emph{stochastic} part is due to averaging over some series of samples, and the fact that these samples fit into the concept of RVE (Representative Volume Element) in order to reduce the variance effect. 

In this article, we present a novel method for computation of effective electric properties of composites -- it is based on the analysis 
of the connectivity graph (and the respective adjacency matrix) for each sample of a composite material. We describe how this matrix is constructed 
in order to take into account complex microscopic geometry. We also explain what we mean by homogenization procedure for electrical conductivity, and how the constructed matrix is related to the problem. The developed method is applied to a test study of the influence of micromorphology 
of composites materials on their conductivity.


%  \subclass{MSC code1 \and MSC code2 \and more}
\end{abstract}

\maketitle

\section{Introduction / motivation}
\label{sec:intro}

Composite materials and more generally heterogeneous media appear in various real-life situations.
The latter represent the matter from microscopic scale to perfectly observable macro objects, while the former
play an important industrial role in the last decades.
The incredibly fast development of composite materials in most of the branches of modern technology is due to their
exceptional properties in comparison with pure materials.
The process of conception and production of composites is however far from being simple:
while a lot of tests are needed to be able to validate the technology before
going to industrial applications, the experimental work is still expensive
and not easy to carry out. All this results in a strong demand for efficient mathematical modelling
techniques that can be applied to predict and optimize effective properties of composite materials.
Moreover in the recent years computational facilities have become much more accessible, that resulted
in a great number of works dealing with development of methods and implementation of algorithms for such
studies\footnote{Similar logic goes far beyond the subject of composite materials: one can for example recall the
Nobel Prize 2013 in chemistry, that was given more or less for advances in computer simulations.}.
Out of all possible approaches to the problem we focus on the method which is usually called \emph{stochastic homogenization}.

The key idea of \emph{homogenization} as a modelling technique\footnote{not to be confused with an industrial process}
is to be able to estimate measurable macroscopic parameters of a composite material (or heterogeneous medium)
starting from its microscopic geometry. In a sense, one computes the parameters of an equivalent homogeneous material
that can ``replace'' the studied heterogeneous medium in the experiment.
Various techniques based on different mathematical and algorithmic approaches have been proposed.
In what follows, we will see that it is very important to carefully choose suitable methods, and
this choice is greatly influenced by the problem one is solving or by a property one is studying,
and also by availability and type of the input data.
In our previous papers we have observed for example, that the methods based on Fast Fourier Transform proposed initially in \cite{moul-suq, monchiet}
for 2D problems, are perfectly suitable for studying
elastic properties in 3D (\cite{VSPD}), but they have to be modified to be applied to thermal phenomena (\cite{SVDP2, SVDP3}).
It became clear from the previous studies (\cite{SVDP4}) that they are practically not efficient to estimate electrical conductivity, which is the main motivation
for the method we describe in this paper.

The \emph{stochastic} part in the name of the method is due to the fact that each given sample
of a medium does not capture exactly the behavior of the composite material. One is thus
forced to average the results on a reasonable sampling with some common properties.
The idea is coming from a very classical approach of Monte Carlo simulations.
For our purposes, it means that an efficient algorithm of generation of such samples should be implemented.
We profit from the flexibility of the methods we have developped in \cite{VDP, SVDP} to use them
in the analysis of electrical conductivity.

Our interest in the subject is also motivated by some very concrete applications related to aeronautic
equipment, a project carried out in close collaboration with industrial partners.
Some part of this contribution will thus be devoted to explaining in what way the techniques can indeed be
applied and to potential difficulties that arise in ``true-to-life'' situations, and to possible ways out.

The main goal of this paper is to provide a tool that can be applied ``out of the box'' in order to
study effective properties of heterogeneous media or composite materials. Therefore we present the ``building blocks''
for such a tool, that are sufficiently flexible and can be then ``assembled'' depending on the available input  data
and the problem under consideration.
 For each method that is discussed here, we give ideas of some mathematical
 background, that is necessary for understand and application, we do not however go into much details
 not to overload the presentation.

To sum it up, the text is organized as follows:
In the next section (\ref{sec:stoch}) we briefly recall the
tools we have developped for generating samples of composite materials/heterogeneous media with rather complex microscopic geometry;
we explain how they can be easily adapted to cover the situations arising typically in the studies of electrical conductivity.
In the section \ref{sec:hom} the novel (deterministic) homogenization procedure is presented -- it is based on the analysis of the connectivity graph
of a material and is perfectly suitable for the type of samples we produce. The last section (\ref{sec:appl}) is devoted to application
of the developped method to some test examples, it also contains some discussion about its adaptation to industrial problems.

\section{Stochastic part -- sample generation}  \label{sec:stoch}

A composite material, or what is the same in our analysis, a heterogeneous medium, is a material composed of several
constituents often called phases. Typically one thinks about some ``matrix'', say made of a polymer, into which inclusions
of a different nature are inserted. For electrical conductivity the matrix would be made of an insulator \modif{(or a material with rather low conductivity)} and
inclusions would be microparticles chosen to conduct electricity \modif{(typically containing a metal/alloy or some other material with high conductivity)}. It is intuitively clear (and also confirmed by a lot of studies)
that effective properties of composite materials strongly depend on the volume fraction of inclusions, but sometimes even
in a more pronounced way on the \emph{morphology} of the material, i.e. microgeometry and repartition of inclusions.
For modelling, it is thus important to be able to generate samples that are close enough to the actual microstructure of the material.
But since we talk about stochastic homogenization, i.e. averaging the results over a big series of tests, this process should not
take too much time. Not going much into details, let us just say that the usual way out is to replace the
inclusions by simple geometric objects and handle their repartition inside the sample. And certainly a lot of work has been done
in this direction (see \cite{VDP, SVDP} and references therein).

\begin{figure}[ht]
\centering
\subfigure[\, Non-overlapping inclusions]{
 \includegraphics[width=0.45\linewidth]{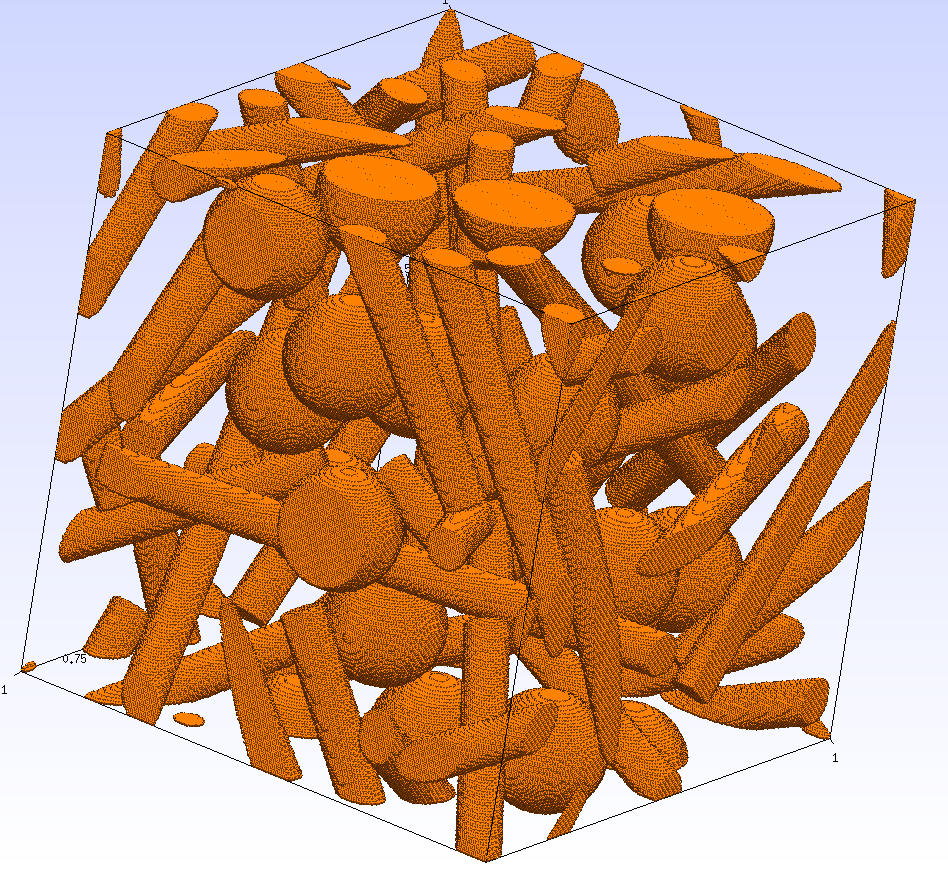}
}
\subfigure[\, Highlighted zone corresponds to overlaps]{
 \includegraphics[width=0.45\linewidth]{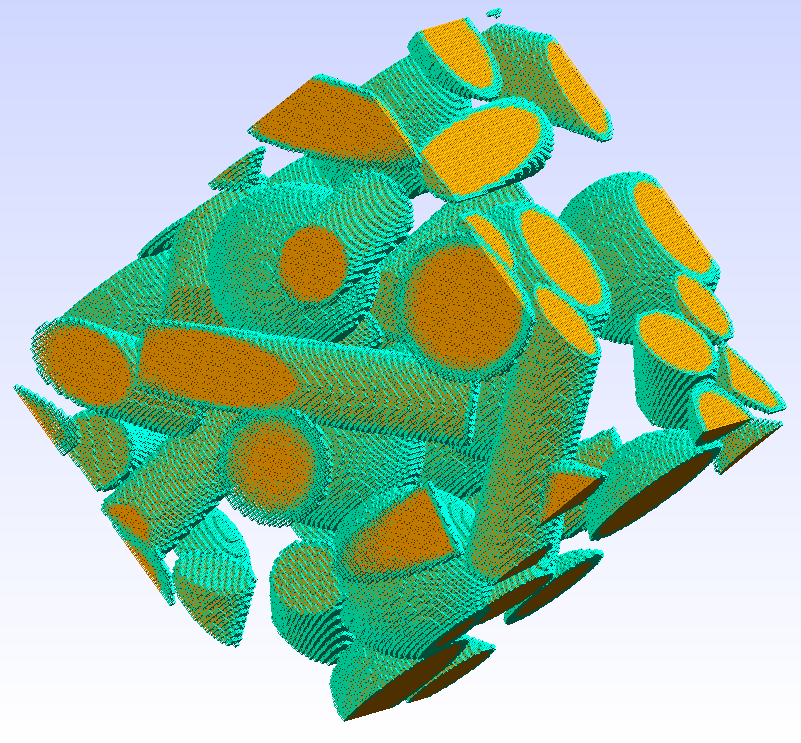}
 }
\caption{\label{fig:rve} 3D view of an RVE: spherical and cylindrical inclusions, periodic boundary conditions.
}
 \end{figure}

The concrete tool we have at hand permits to place spherical and cylindrical inclusions into the matrix, that already
gives a very rich ``true-to-life'' microscopic geometry. The key feature that we need throughout all the studies is the ability to
control contacts and overlaps of inclusions. In the previous studies (elastic and thermal) we were rather focusing on non-overlapping
or slightly overlapping inclusions -- the machinery had been developped mostly for that.
Let us be slightly more specific about the algorithm. There are essentially two methods how to generate samples
with non-intersecting inclusions: RSA- and MD-based ones. RSA (\cite{rsa1}) stands for ``random sequential addition'' -- the inclusions
are generated one after another, and only those that do not intersect with previously generated ones are kept.
MD (\cite{LSA}) stands for ``molecular dynamics'' -- all the inclusions are generated simultaneously, and then repulsive forces are introduced
to position them correctly. Both algorithms have been adapted to the case of spherical and cylindrical inclusions and implemented
 in \cite{VDP}. The input of the algorithms consists of the desired volume fraction, the respective number of spheres and cylinders and their
 geometric parameters. 
 The output is a list of non-intersecting inclusions with coordinates of their centers and
 angles responsible for orientation, this can be used directly in computations or
 converted to a 3D image like figure \ref{fig:rve}.(a).

 Let us note that the numbers should be chosen in order to fit the concept of a Representative
 Volume Element (RVE, \cite{kanit}): the sample should be large enough to effectively capture the properties of the material, like isotropy for example,
 but sufficiently small to make the computation doable.  \modif{We have performed a detailed study of the effective elastic properties of composite materials in \cite{VSPD} and it included also some tests to choose the size of the RVE. Let us just mention here that if the volume fraction and the proportion of each type of inclusions are fixed, the choice of the size of the RVE is governed by the number of inclusions. The usual way to determine this number is to perform some test computations increasing it, and stop when the computed values are no longer affected.}

Though perfectly suitable for computations of elastic properties and in some cases of thermal ones, the 
generation procedure we have just described, clearly does not reflect the way how materials with  good electrical conductivity are produced: the inclusions there should
overlap or at least have some contacts to ensure continuity in the electric network they form.
To cure this we use a simple trick to ``puff up'' the inclusions, i.e. increase their linear sizes, figure \ref{fig:rve}.(b). This does not spoil any isotropy,
but creates eventual contacts and overlaps.
The result is then a network of inclusions with eventual pairwise overlaps.
The fact that this depicts rather well
the real conducting composite materials used in industry may sound surprising, but actually it should not, since we are
more or less modelling the actual process of their production. The first step of generating non-intersecting
inclusions corresponds to placing them into the matrix, while ``puffing up'' in a simplified way
mimics various external forces applied to the samples to ensure contacts. This qualitative logic is in some sense inspired by \cite{GSS},
where some analysis of differential equations related to molecular dynamics is performed.

\section{Homogenization for electrical conductivity} \label{sec:hom}
After applying the method discussed in the previous section, we have a series of samples for each set of macroscopic parameters, let us now turn to the
procedure for computing the effective electrical conductivity of those.

Certainly the best approach would be to honestly consider the Maxwell's equations:
\begin{eqnarray}
    \mathrm{div} \; \mathbf{D} = \rho_V, &\quad&    \mathrm{div} \; \mathbf{B} = 0, \nonumber \\
    \mathrm{rot} \; \mathbf{E} = -\frac{\partial \mathbf{B}}{\partial t},  &\quad&
    \mathrm{rot} \; \mathbf{H} = \frac{\partial \mathbf{D}}{\partial t}  + \mathbf{J}, \nonumber
\end{eqnarray}
where $\mathbf D$ is the electric flux, $\rho_V$ -- charge density, $\mathbf{B}$ -- magnetic flux,
$\mathbf{E}$ --  electric field, $\mathbf{H}$ --  magnetic field, $\mathbf{J}$ -- electric current density.
One should consider these equations %three times
with different boundary conditions in order to recover the whole conductivity tensor.
It is convenient to rewrite the equations only in terms of fields and not fluxes, using that
$\mathbf{D} = \varepsilon \mathbf{B}$, $\mathbf{B} = \mu \mathbf{H}$, with
$\varepsilon$ -- permittivity of the material, $\mu$ its permiability; and
$\mathbf{J} = \sigma \mathbf{E}$ (Ohm's law) with $\sigma$ -- the conductivity.
This gives:
\begin{eqnarray}
    \mathrm{div} \; \mathbf{E} = \frac{\rho_V}{\varepsilon}, &\quad&    \mathrm{div} \; \mathbf{H} = 0, \nonumber \\
    \mathrm{rot} \; \mathbf{E} = -\mu \frac{\partial \mathbf{H}}{\partial t}, &\quad&
    \mathrm{rot} \; \mathbf{H} = \varepsilon \frac{\partial \mathbf{E}}{\partial t}  + \sigma \mathbf{E}. \nonumber
\end{eqnarray}
Like this one sees the dependence on parameters coming from the medium, like $\varepsilon, \mu$ and $\sigma$.
And in our case these parameters do depend on spatial coordinates, since they are different for all the phases of the material.

In fact the problem
can be significantly simplified since we are interested in a quasistationary regime without any charges inside
the medium. Half of the equations thus become trivial, while the other half reduces precisely to
the Fourier's law in thermodynamics, with the temperature $\theta$ being the analogue of the electric potential, 
the heat flux $\boldsymbol{\phi}$ -- of the current, \modif{and $L(\boldsymbol{x})$ -- the conductivity tensor}:
\begin{equation} \label{fourier}
\boldsymbol{\phi}(\boldsymbol{x})=-L(\boldsymbol{x})\: \nabla \theta(\boldsymbol{x}), \qquad
\mathrm{div}(\boldsymbol{\phi}(\boldsymbol{x})) = 0.
\end{equation}
This permits also not to worry about well-posedness of the problem.
One is thus tempted to apply the same FFT-based methods or use  Finite Elements, as in the studies of thermal properties.
Unfortunately, both approaches present serious difficulties.
It has been observed in the original papers that the FFT-based methods are very sensitive to the contrast (ratio of conductivity)
between the matrix and the inclusions, which is very high in the electric case -- this results in a drastically slow convergence of
the method. As for the FEM, the problem comes simply from the size of the mesh needed to capture microscopic geometry.
We are thus brought to the need of a different method that we describe below.

\modif{\subsection*{Description of the method}}
The key idea is to study the structure of the network of inclusions that we have obtained in the previous section:
consider the inclusions as nodes of an electric circuit, and say that those couples that have an overlap
are connected.
One thus obtains a connectivity graph of a sample with vertexes corresponding to inclusions and edges to
contacts. Morally this means that we consider the inclusions made of a conducting material and the matrix is a perfect insulator.
Already with this approach one can say something about conductivity and even more, see for example \cite{Luck}, where networks of randomly distributed resistors are studied.
But the information about connectivity is certainly not enough: we would like not only to say if the
sample is conducting current or not, but also give a numerical estimate of this conductivity. It means that we should assign
some resistance to each edge of the constructed graph. With this information we can then apply the Kirchhoff's laws:
for any internal vertex the sum of currents vanishes, and for any edge the classical Ohm's law holds. The result will be the total current passing through the circuit -- after some renormalization it gives precisely the desired conductivity.
This idea is used very often for modelling complex systems, like traffic flow (\cite{book}), and it is simple enough to be very efficient.
Paradoxically, we use Kirchhoff's laws to actually compute the current, so
``sometimes an electric circuit is just... an electric circuit''...

\begin{figure}[ht]
\centering
\subfigure{ \includegraphics[width=0.45\linewidth]{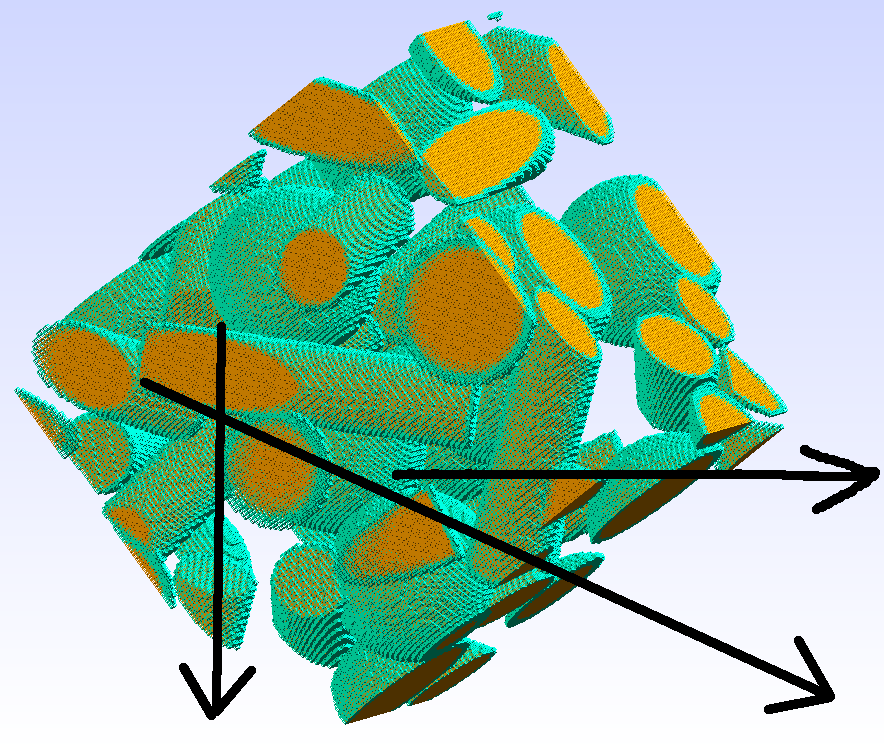} }
\subfigure{ \includegraphics[width=0.45\linewidth]{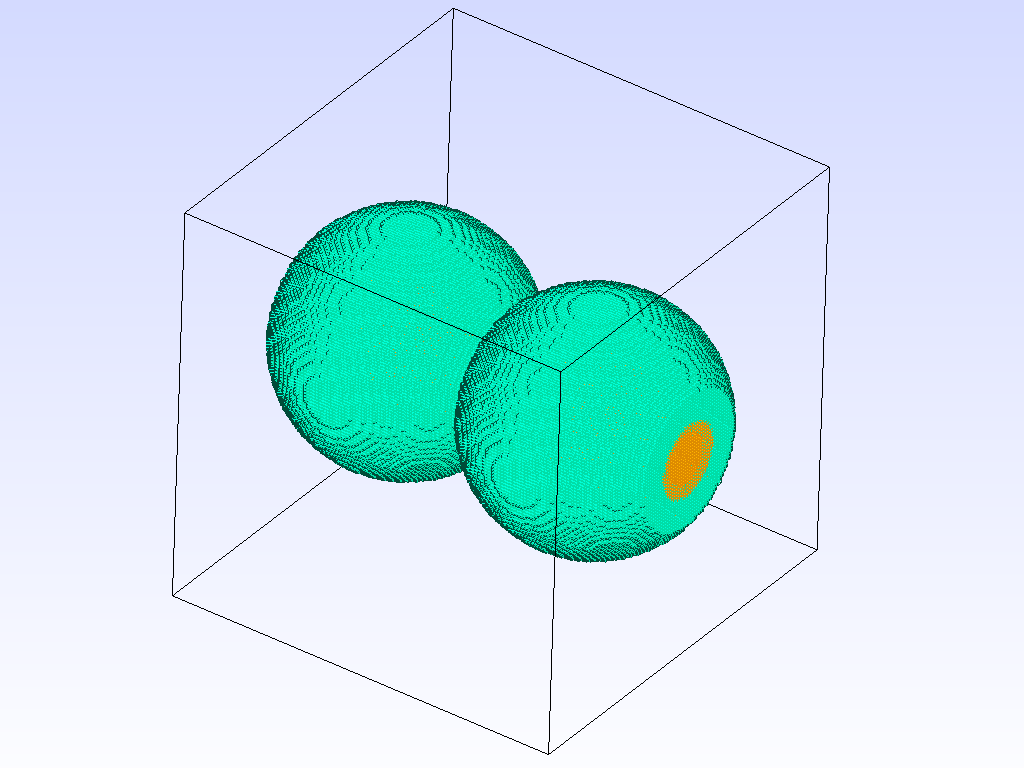} } \newline
\subfigure{ \includegraphics[width=0.45\linewidth]{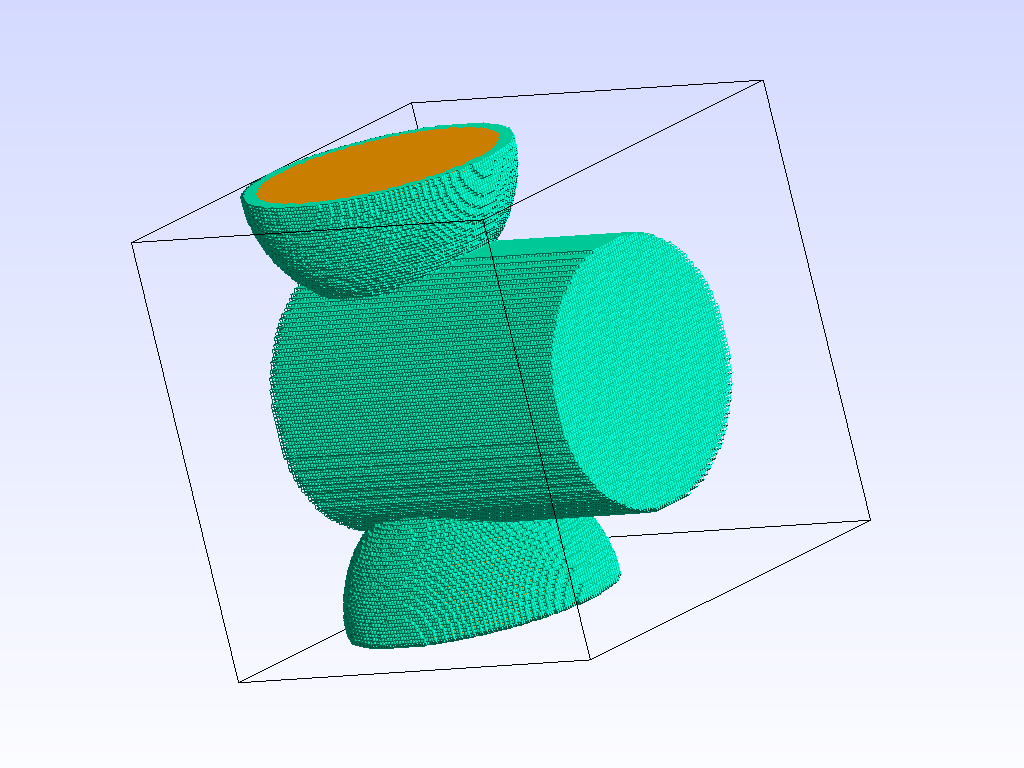} }
\subfigure{ \includegraphics[width=0.45\linewidth]{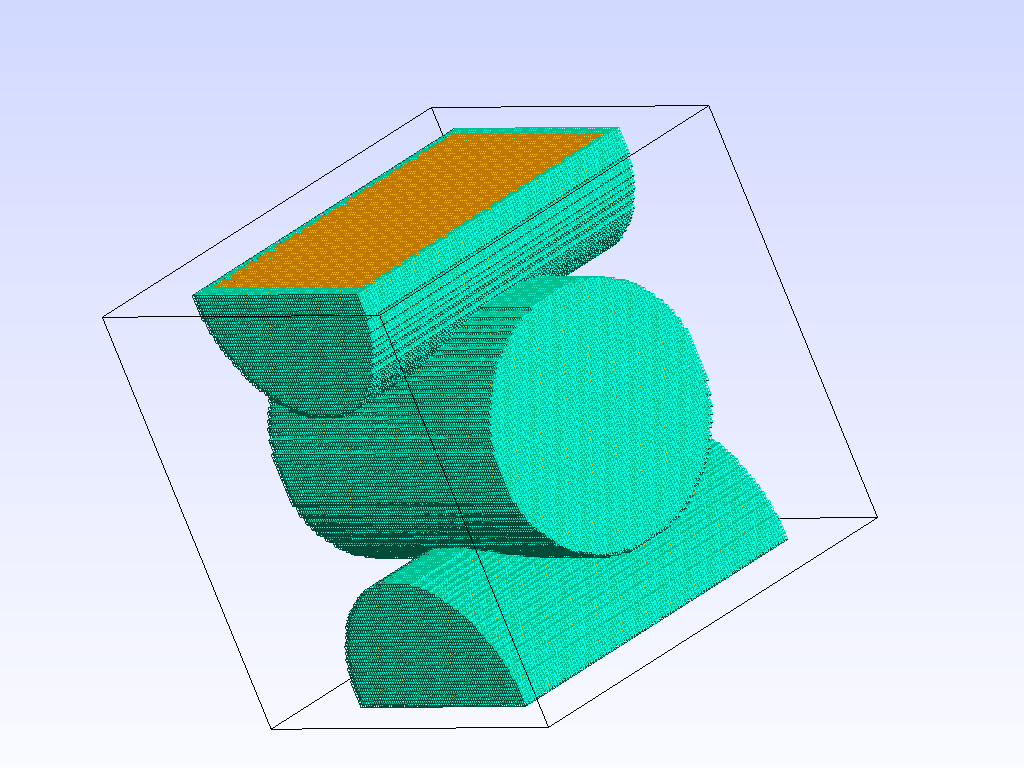} }
\caption{\label{fig:overlaps} Three types of overlaps in a sample.}
 \end{figure}

A crucial detail which remains to be discussed, is how to assign the values of resistance to edges -- and this is a very important question, since complex geometry is taken into account precisely here.
We profit again from the MD-based method of generating the samples
described in \cite{VDP}, namely from the table 1 there, which gives the expressions for repulsive forces
between overlapping inclusions that depend on the geometry of overlaps.
These forces, being rescaled appropriately, will characterize the resistance of the respective edge.
By construction, these forces have elastic nature, so they give a good approximation of the law defining the resistance.
\modif{By rescaling we mean selection of the constant factors in the formulas, since in the original setting they corresponded to mechanical forces and now we suppose the conductance (i.e. the inverse of resistance) to be proportional to those.} 
For the law we consider, one just needs to fix the global normalization for each type of contacts.
This is done by comparing the MD forces with the actual conductivity of the sample computed by FEM from Maxwell's equations,
which is doable for simple geometries. \modif{As one can see from figure \ref{fig:overlaps}, there are essentially three interesting types of overlaps: sphere-sphere, sphere-cylinder and cylinder-cylinder. For all of them small samples containing only a couple of inclusions are considered -- those can be pixelized/meshed at some reasonably small resolution and thus are suitable for FEM. This step might be time consuming, but it is done only once
to calibrate the model. On the other hand, one computes their conductivity using the electric circuit approach: the circuit would then contain just a couple of nodes, where one would adjust the constants to fit the FEM computation. Let us stress that this step is important, since it guarantees good agreement with the FEM that is often considered to be the second (after experimental tests) reliability criterion.  
}
 
\modif{
Certainly, one can consider more involved ways to model the interaction, for instance consider some non-linear dependence between the conductance and the MD force. One then would need to approximate that dependence by some function which will be reconstructed from the simple geometry pairwise overlaps. And again, that would be a potentially time consuming step, but done only once, thus this will not change the global strategy.}
\newpage 

\begin{figure}[ht]
\centering
\subfigure{ \includegraphics[width=0.4\linewidth]{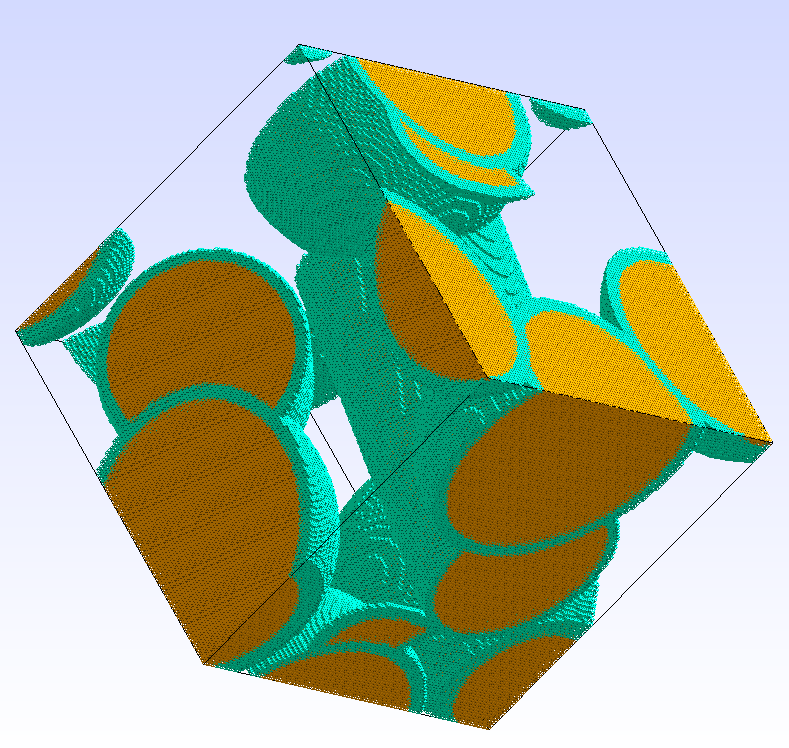} }
\subfigure{  \includegraphics[width=0.56\linewidth]{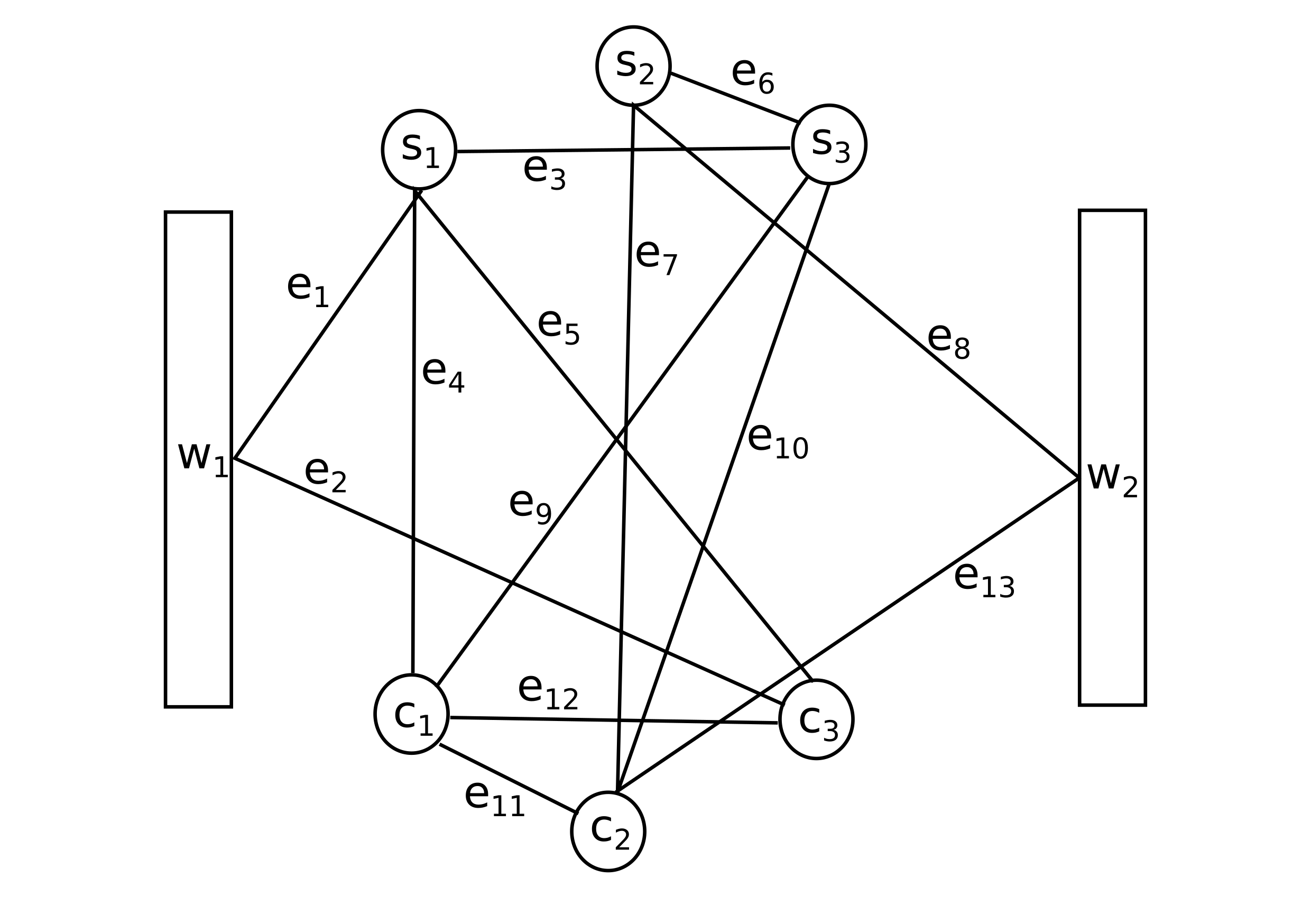}}
\caption{\label{fig:perc} A percolating sample and the corresponding graph.}
 \end{figure}

\modif{\subsection*{Homogenization example}}
 Let us consider a simple example -- a sample with a small number of inclusions
 (figure \ref{fig:perc} -- left). It is of course out of the VER notion, we present it only to illustrate the computation procedure. 
 One can see clearly that this sample is percolating, i.e. there is a path of inclusions
 connecting two boundaries, so even with a small volume fraction of inclusions it does conduct current.
 \modif{
Let us now construct the corresponding graph (figure \ref{fig:perc} -- right ) and its adjacency matrix. As described above all the inclusions
are represented by vertexes: 3 for spheres and 3 for cylinders, we denote them $s_i$ ($i = 1, 2, 3$) and  $c_j$ ($ j = 1, 2, 3$) respectively. We add to them two vertexes that correspond to boundaries: $w_1$ and $w_2$. If an inclusion touches the boundary or another inclusion we draw an edge $e_k$ between the corresponding vertexes  and assign a weight $c_k$ to it equal to the conductance computed from the MD procedure described above. The (weighted) adjacency matrix is constructed by putting the values $c_k$ on the intersection of lines and columns corresponding to the vertexes that define the edge $e_k$,}
then
the resulting $8 \times 8$ matrix has the following form:
$$
  Adj = \left(
  \begin{BMAT}{cccccccc}{cccccccc}
    0   & \framebox{0.3} & 0 & 0 & 0 & 0  & \framebox{0.79} &  0  \\
     \framebox{0.3} & 0  & 0 & \framebox{0.05} & \framebox{0.31} & 0 & \framebox{0.41} & 0 \\
    0 & 0 & 0 & \framebox{0.05} & 0 & \framebox{0.41} & 0 & \framebox{0.38} \\
    0 & \framebox{0.05} & \framebox{0.05} & 0 & \framebox{0.43} & \framebox{0.31} & 0 & 0 \\
    0 & \framebox{0.31} & 0 & \framebox{0.43} & 0 & \framebox{0.01} & \framebox{0.14} & 0 \\
    0 & 0 & \framebox{0.41} & \framebox{0.31} & \framebox{0.01} & 0 & 0 & \framebox{0.28} \\
    \framebox{0.79} & \framebox{0.41} & 0 & 0 & \framebox{0.14} & 0 & 0 & 0 \\
    0 & 0 & \framebox{0.38} & 0 & 0 & \framebox{0.28}& 0 & 0
    \addpath{(0,3,1)ddruuuuuurrrrrrulllllllddddd}
    \addpath{(1,2,1)ddrrrrrrruuuuuuulddddddllllll}
  \end{BMAT}
  \right)
$$
\modif{This corresponds to the ordering of vertexes: $w_1, s_1, s_2, s_3, c_1, c_2, c_3, w_2$, and
the delimited upper-left and lower-right corners represent the opposite boundaries $w_1$ and $w_2$.}
The matrix is symmetric, and only off-diagonal values are important. The boxed non-zero numbers come from
the construction described above, there are precisely $13$ of them, since there are $13$ non-trivial contacts, \modif{and if one looks only at the upper triangle they are ordered line by line in the same way as 
$c_k$'s.

In principal this data would be already enough to solve the problem, but in order to make the implementation more transparent let us introduce one more graph-related object -- the incidence matrix $A$. The lines of it are indexed by the vertexes of the graph and the columns by the edges. The entries are $1$ if the edge comes out or goes into a given vertex, and $0$ otherwise. 
In our example for the same ordering of vertexes as before it reads:}
$$
A = \left(
  \begin{BMAT}{ccccccccccccc}{ccccccccccccc}
  1&1&0&0&0&0&0&0&0&0&0&0&0\\
  1&0&1&1&1&0&0&0&0&0&0&0&0\\
  0&0&0&0&0&1&1&1&0&0&0&0&0\\
  0&0&1&0&0&1&0&0&1&1&0&0&0\\
  0&0&0&1&0&0&0&0&1&0&1&1&0\\
  0&0&0&0&0&0&1&0&0&1&1&0&1\\
  0&1&0&0&1&0&0&0&0&0&0&1&0\\
  0&0&0&0&0&0&0&1&0&0&0&0&1
  \end{BMAT}
  \right)
$$
\modif{This matrix can be read off from $Adj$ or constructed directly, provided the number of edges is known. If one considers the vector $c = (c_1, c_2, \dots)$, then the weighted degree matrix 
$D := diag(A \cdot c)$ can be constructed, as well as the conductance matrix $C:=diag(c)$. 
One then recovers the adjacency matrix $Adj = A \cdot C \cdot A^T - D$.

But the most important application of $A$ is not the identity above. Actually, it carries the information about the structure of the linear system that will be used to compute the effective conductivity of the sample. 
Let us enumerate the unknowns, which split into two types: the values of the electric potential at each vertex ($8$ of them here, denoted $u_1, u_2, \dots$),
and the current through each edge ($13$ of them, denoted $I_1, I_2, \dots$). 
One easily notices that the columns of $A$ correspond to currents, namely they encode the Ohm's law for each edge: $I_k = c_k \cdot (u_{in(k)} - u_{out(k)})$, where the indexes for $u$ are precisely the lines with non-zero entries for the $k$-th column; this gives 13 linear conditions.
The lines (all but the first and the last ones) correspond to the Kirchhoff's law for the internal nodes:
$\sum\limits_{A_{ik \neq 0}} I_k = 0$; this gives 6 more linear conditions. 
And we impose some values of potential on the boundary, say $0$ and $1$, which complete the linear system. Clearly this counting of unknowns and equations remains valid for any graph. 
Then it is enough to solve
the obtained system, or better to say compute the total current through the circuit, to obtain the effective resistance of the sample
in the direction between the two boundaries. Here again the structure of the matrix $A$ is helpful: this total current is the sum of outgoing currents from the first boundary (indexed in the first line of $A$)
or of incoming currents to the second boundary (the last line of $A$ respectively).
In this concrete example the computed value is approximately $0.11$.

This is the scheme of the \emph{homogenization procedure} we suggest. 
To complete the picture, one needs to perform similar computations for all the couples of boundaries of the domain to recover the whole conductivity tensor $L$ from the equation (\ref{fourier}).
The opposite faces of the cube will correspond to diagonal and the neighboring to off-diagonal entries. For these latter ones it is important to treat adequately the inclusions that are in contact with both faces: a possible reasonable choice is to consider only the central part of the boundary, or even just the central point. 
}

A couple of technical remarks are in place here.
First, we have reduced the homogenization problem to solving a linear system. It may be rather large for ``by hand'' solution, but it
is very small if one compares with those that arise in for example Finite Elements. This means that
basically any available solver would do the job, one just need to pay attention to possible degeneracies.
We mean the following: the system is never overdetermined and it is consistent by construction, but it can be degenerate.
This is easy to understand from the physics of the problem: the graph we construct from the network of inclusions need not be
connected (imagine an isolated inclusion inside the matrix) then to the parts not connected to boundaries
one can assign any value of the electric potential. But this is the only freedom in the solution, so the result of the
homogenization procedure is well defined.
Second, in contrast to our toy example, already the adjacency matrix in the typical situation is rather sparse, and this is even
more pronounced for the final linear system. One can thus profit from this knowledge to
choose appropriate solvers and optimize even more the computation time. Alternatively, one can remember that the system was coming from graphs, and
profit from various packages available for their analysis.

\section{Test examples, applications, and discussion} \label{sec:appl}

Let us now put the procedures described in the previous sections together and see what happens
in real stochastic homogenization computations.\\[-2.5em]

\subsection{Test examples}

\begin{figure}[ht]
\centering
\subfigure[\, Cylinders' length/radius = 3.]{ \includegraphics[trim = 2.2cm 0cm 2.2cm 0cm, clip, width=0.45\linewidth]{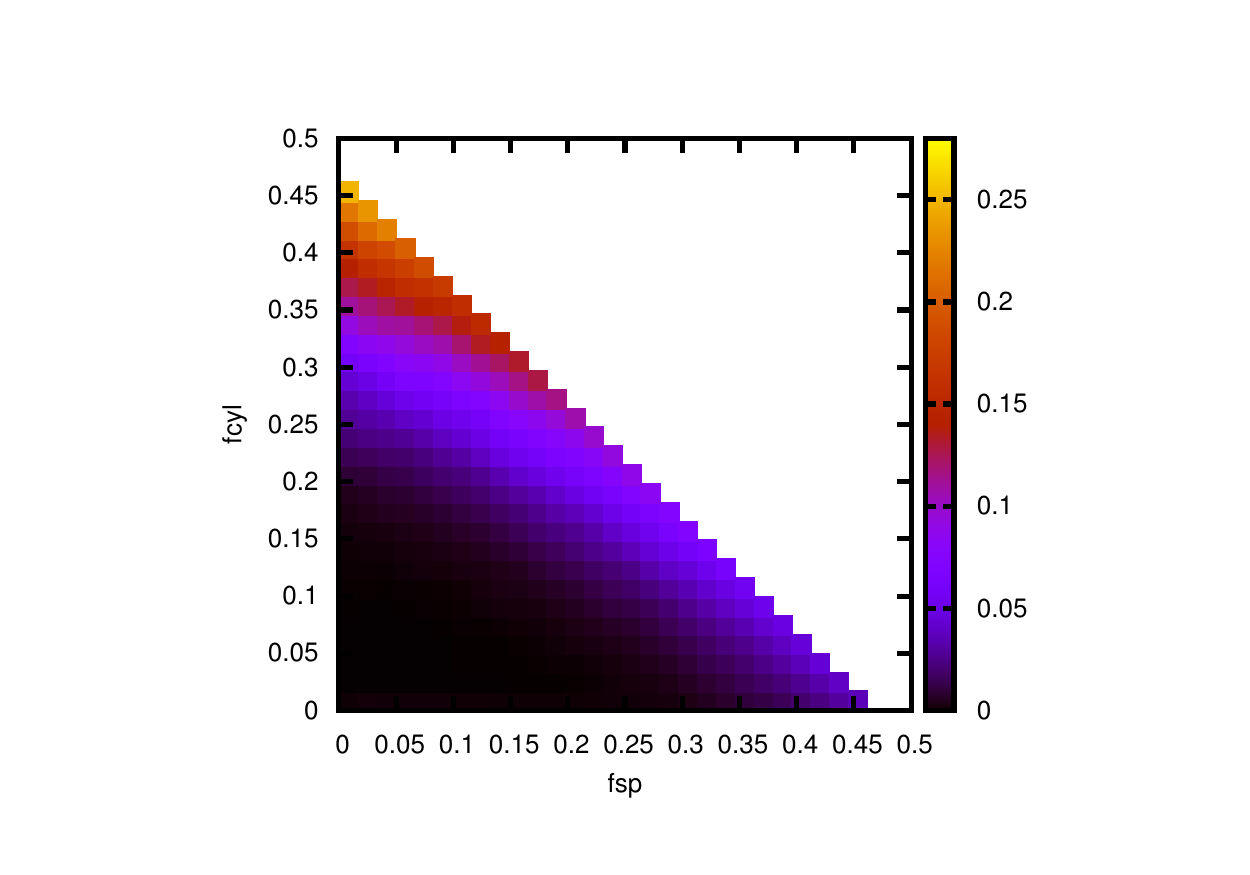} }
\subfigure[\, Cylinders' length/radius = 5.]{ \includegraphics[trim = 2.2cm 0cm 2.2cm 0cm, clip, width=0.45\linewidth]{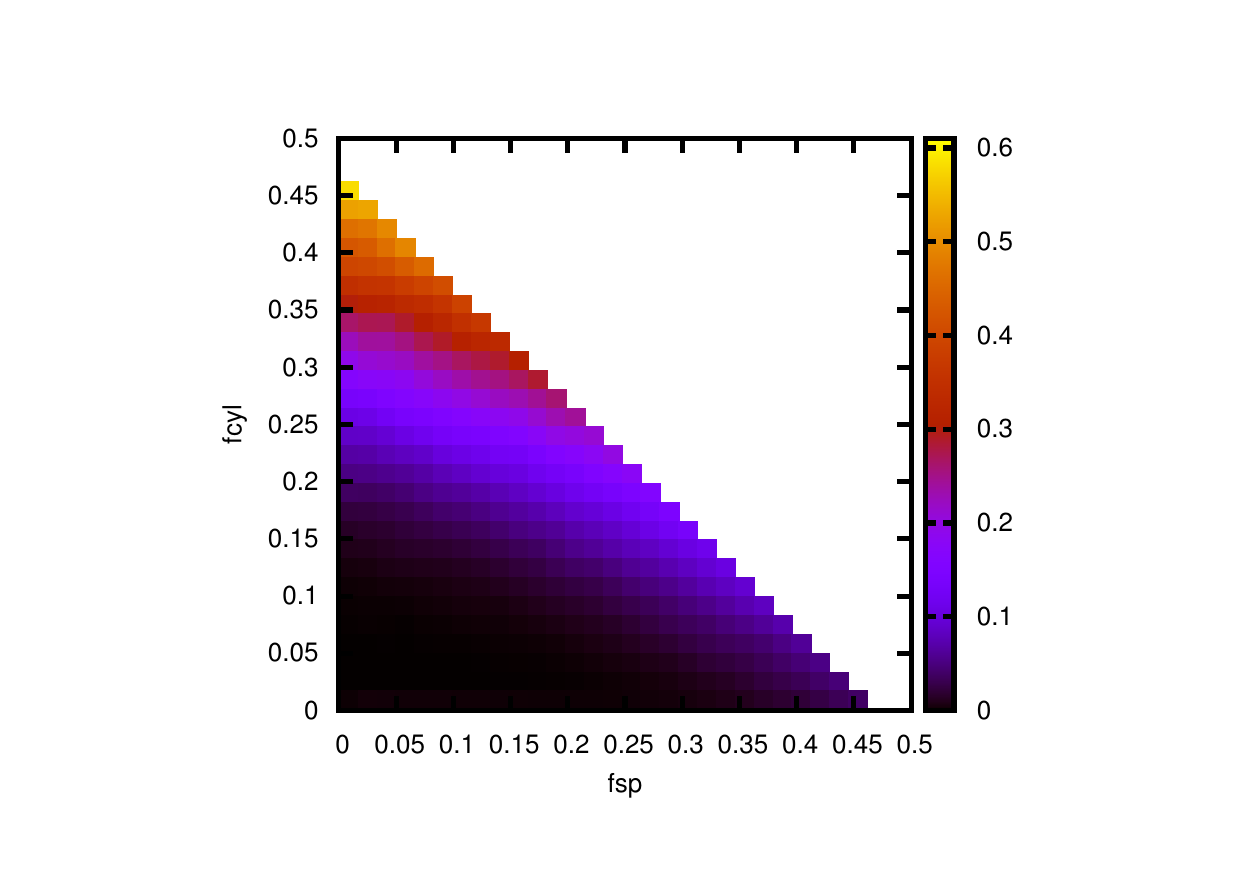} }
\caption{\label{fig:comp} Dependence of conductivity on the repartition of  volume fraction of inclusions. \modif{The values are given in dimensionless units, the value of $1$ corresponding to a sample fully made of the conductive material.}}
 \end{figure}

We consider the samples containing a mixture of spherical and cylindrical inclusions, and study the dependence
of their electrical conductivity on the volume fraction of each type of inclusions. Below (figure \ref{fig:comp})
we plot this value for two types of tests, for different aspect ratios (length/radius) of cylinders, the value is
averaged over a series of samples with the same macroscopic parameters, but different microgeometry. \modif{The output is given in dimensionless units, where the value of $1$ corresponds to maximal conductivity, i.e. the sample made of pure conducting material, or equivalently the volume fraction of inclusions being $100\%$.}
One sees that for longer cylinders the conductivity of samples increases in the average.
Certainly this is rather a consistency test for the method we suggest than a real systematic study
of the effects of morphology of inclusions on the effective properties of materials, that we intend to
do in a separate article.
The main message of this paper is precisely that using rather simple and accessible tools and computations that run in fractions of
a second, one can already obtain
convincing results.
\\[-2.5em]

 \subsection{Industrial applications}
As we have already mentioned, behind this study there is a concrete applied problem related to
estimation and improvement of effective properties of composite materials.
In real-life applications one does not work with precisely the same samples that we have discussed in
section \ref{sec:stoch}, they do look like figure \ref{fig:rve}, but have a completely different origin.
Usually one obtains this picture not by pixelizing the vector data of the set of inclusions, but the
other way around -- for example from tomography or microscopy. Then image processing
algorithms are used to segment the image, i.e. distinguish the material that is represented by each voxel.
Even though modern algorithms permit to make this distinction if a given voxel belongs to the matrix or to inclusions
with rather good certainty, the information to which inclusion precisely it belongs is not available directly.

But we are still able to handle the situation with our approach. In a sense, the geometric part is even
simpler in this case: one considers voxels corresponding to inclusions
as independent particles, and the connectivity graph is just the information about neighbouring voxels.
As for the forces assigned to edges, now one needs to consider three cases of voxels that meet by a face, by an edge or by a vertex,
and adjust the constants appropriately.
The price one pays is certainly the size of the resulting linear system,
so the observation about sparsity becomes extremely important.
  This looks like a ``baby version'' of Finite Elements, but it is still
much simpler both from mathematical and from algorithmic points of view.
Moreover, the computation being local, the algorithm should be suitable for parallelization. \\[-2.5em]

\subsection{Discussion}

We have described a procedure permitting to compute effective electric properties of composite materials
using very efficient and accessible tools. Since the required computational resources are very modest it
is appropriate for stochastic homogenization. This provides a tool both for analytical studies and industrial applications.
However there is a couple of questions we are curios about and intend to address them in the nearest future.
First, as mentioned before, the test problems and some research ones do not need huge computational resources, while
industrial applications eventually do. In what we have discussed above,  the solution of the linear system was (theoretically)
exact, but in real life this might be not needed.
We think that the link with graphs in the context is very promising and can provide efficient and reliable ways to handle the problem.
Second, by construction our method is in good agreement with FEM, since the latter one is implicitly used.
It means that at least for simple geometry of samples, the obtained solutions reproduce well the ones obtained
from Maxwell's equations.
It would be interesting to observe and eventually prove the same thing for the FFT-based methods.

\textbf{Acknowledgements.}

This work has been supported by the ACCEA project selected by the ``Fonds Unique Interminist\'eriel
(FUI) 15 (18/03/2013)'' program.
The research of V.S. was also supported by the Fonds National de la Recherche, Luxembourg, project F1R-MTH-AFR-080000.

\end{document}